\documentclass[12pt]{article}
\input{epsf.tex}
\begin{document}
\begin{center}
{\bf PARAMETERS OF THE BEST APPROXIMATION FOR DISTRIBUTION OF THE
REDUCED NEUTRON WIDTHS
. ACTINIDES}
\\\end{center}
 \begin{center}
{\bf  A.M. Sukhovoj, V.A. Khitrov}\\
\end{center}
\begin{center}
{\it Joint Institute for Nuclear Research, Dubna, Russia}
\end{center}

\begin{abstract}
The data of ENDF/B-VII library on $\Gamma^0_n$ ($\Gamma^1_n$) for nuclei
$^{231}$Pa, $^{232}$Th, $^{233,234,235,236,238}$U, $^{237}$Np,
$^{239,240,241,242}$Pu, $^{241,243}$Am and $^{243}$Cm
(including p-resonances of $^{232}$Th, $^{238}$U, $^{239}$Pu) in form
of cumulative sums in function on $\Gamma^0_n/<\Gamma^0_n>$ were
approximated by variable number $K$ of partial items  $(1 \le K \le 4)$.
Parameters of approximation -- mean value of neutron amplitude,
its dispersion and portion of contribution of part of widths of distribution
number $K$ in their total sum. 
The problems of their determination from distributions of different
number of squares of normally distributed random values with variable
threshold of loss of some part of the lowest $\Gamma^0_n$ values were studied.  

It was obtained for some part of neutron resonances that their mean
amplitudes can considerably differ from zero value, and dispersions --
from $<\Gamma^0_n>$.
And it is worth while to perform any quantitative analysis of
distributions $\Gamma^0_n$ by means of comparison of different model
notions with obligatory estimation of random dispersion of the desired
parameters.
\end{abstract}

\section{Introduction}\hspace*{16pt}

The experimental data on reduced neutron widths $\Gamma^0_n$ ($\Gamma^1_n$)
of s- or p-resonances potentially contain diversiform information.
In particular, of interest are the data on their real density $\rho$
and on few-quasi-particle components of wave function of high-lying
level \cite{PEPAN-1972}.
Restoring of this information from the data of any experiment
on the neutron time of flight method cannot be made without
the use of hypothesis on form of the frequency distribution of neutron
widths. Porter and Thomas first assumed \cite{PT}, on the grounds
of large fluctuations of $\Gamma^0_n$, that it corresponds to
$\chi^2$-distribution with one degree of freedom.
In the other words -- to distribution of squares of normally
distributed random values $\xi$ with mathematics expectation
$M(\xi)=0$ and dispersion $D(\xi)=1$.
These conditions are realized to higher or lower extent in case
when wave function of neutron resonance contains many small items
of different sign which determine the value $\Gamma^0_n$.
Within frameworks of notions of quasi-particle-phonon nuclear model,
this means that fragmentation of any nuclear states like $m$
quasi-particles $\otimes$ $n$ phonons over nuclear levels in
region of neutron resonances (for truth of \cite{PT}) must be very strong.

Theoretical analysis of fragmentation process of nuclear states of
different type over the levels with different excitation energy 
\cite{MalSol} showed that this condition can be not realized in case
of large enough values $m$ and $n$. This conclusion follows and from
results of approximation below $B_n$ \cite{Prep196, PEPAN-2006} of
methodically more precise experimental data for $\rho$ \cite{Meth1,PEPAN-2005}
on density $\rho_n$ of $n$-quasi-particle levels by Strutinsky model
\cite{Strut}. It can be performed at wide enough variation of
assumptions on shape of correlation function of nucleon pair in
heated nucleus and on coefficients of its vibration enhancement.
Nevertheless, results of approximation already gave some notion
on structure of high-lying levels of any nuclei.  

The results  \cite{Prep196, PEPAN-2006} unambiguously show that
at the simplest hypotheses of correlation functions $\delta_n$
of nucleon Cooper pairs in heated nucleus, the number $n$ increases
by 2 quasi-particles with excitation energy interval being some less
than $\Delta E_{\rm ex} \approx 2\delta_0$. Id est, structure of
wave function of highly-excited levels (and, it is not excluded,
of neutron resonances) can cyclically change. These data are enough
for qualitative explanation of change in form of radiative
strength functions as increases mass of nucleus $A$ \cite{Appr-k}.
Id est, there are theoretical and experimental grounds for detailed
and methodically independent analysis of the data on $\Gamma^0_n$.
The primary goal of this analysis -- discovery of possible
deviations of neutron width distribution from the Porter-Thomas
distribution and estimation of reliability extent for their
observation in experiment. 

Reanalysis of the data on gamma-transition intensities from reaction 
$(\overline{n},\gamma)$ \cite{Avres,AllNuc} revealed strong influence
of nuclear structure  on distribution of parameters of radiative widths
of primary gamma-transitions from neutron resonances.
This is an additional argument for independent full-scale analysis of
the data on neutron widths with less quantity of the used model ideas.

\section{Modern status of the problem}\hspace*{16pt}

Direct determination of structure of arbitrary nuclear levels above
some MeV usually is inaccessible for all the known experiments.
Therefore, any information on this account can be derived only from
indirect data (as it was shown in \cite{PEPAN-1972}).
First of all -- from analysis of the results of model approximation
of experimental distributions of $\Gamma^0_n$ ($2g\Gamma^0_n$ --
for resonances with different spins). Numerous problems of analysis
of experimental data of this type are described, for example,
in \cite{Fro83}. Analysis of modern data on parameters of neutron
resonances of $^{238}$U in neutron energy region up to  20 keV is
presented, for example, in \cite{Derr2005}, $^{232}$Th - \cite{2008De20}.

The basis for all the performed earlier analyses is an assumption
that the hypothesis \cite{PT} describes tested set of the data on
$\Gamma^0_n$ with a precision exceeding accuracy of the experiment.
By this it was assumed that in analysis is in some form realized
correct accounting (or exclusion) of experimental distortions of
the data under consideration (omission of weak levels, unresolved
multiplets, admixture of resonances with other orbital
momentum $l$ and so on). Or observed discrepancies of
experiment with distribution \cite{PT} are completely explicable by
enumerated factors. In practice, it is tested up to now only the
hypothesis of deviation of experimental distribution of $\Gamma^0_n$
from the expected theoretical one owing only to deviation
of parameter $\nu$ of theoretical distribution from unit.

Whereas, the form of distribution of $\Gamma^0_n$ strongly depends
on degree of execution of condition of equality to zero of mean
amplitude $A_n$ ($\Gamma^0_n=A^2_n)$ of the tested set of resonances.
It is absolutely impossible also to exclude a possibility that the
experimental data are superposition from $K$ distributions even for
their set with precisely determined spins and orbital momenta of
resonance neutron.
Qualitatively, the possibility $M(A_n) \ne 0$ directly follows from
\cite{MalSol},  $K>1$ -- from \cite{Prep196, PEPAN-2006}.

\section{Necessity in full-scale analysis of experimental data}\hspace*{16pt}

The fitted function in full-scale analysis is the sum of $K$
distributions $P(X)$ of squares of normally distributed random
values with independent variables $X_k$ each.
The desired parameters in compared variants are the most probable
value $b_k$ of amplitude $A=\sqrt{\Gamma_n^0/<\Gamma_n^0>}$,
dispersion $\sigma_k$ and total contribution  $C_k$ of function
number $k$ for the variable

\begin{equation}
X_k=((A_k-b_k)^2)/\sigma^2_k
\end{equation}
in the total experimental sum of widths. Statistically significant 
result $b_k \neq 0$ allows one to state that the neutron resonance
is not completely chaotic system, $\sigma_k < 1$ means, in particular,
that interaction of neutron with non-excited nucleus is the more or
less determinate process.

The proof of the notion that the most precise approximation of
dispersion of widths by several distributions ($K>1$) is not caused
by random fluctuations of $X$, can give new information on nuclear
structure in region $B_n$. First of all -- information on possible
existence of neutron resonances with different structure of their
wave functions and on regions of $\Gamma^0_n$ values, where radiative
strength function (the total gamma-spectrum) have essentially
different form (see, for example \cite{Appr-k}).

Cyclic change in structure of neutron resonances at different
neutron energies (directly following from successive break up of
Cooper nucleon pairs \cite{Prep196, PEPAN-2006}) can stipulate
non-monotonous character of change in density of nuclear excited
levels and above neutron binding energy. This means additional
systematical error of the data on $\rho$ in the most important
for this nuclear-physics parameter point.

As it was obtained by modeling \cite{AMSa}, the values of $b_k$,
$\sigma_k$ and $C_k$ with small statistical error for accumulated
by now sets of neutron widths cannot be get not only for $K>1$,
but also for $K=1$.
Most probably, this circumstance has principle character and appears
itself,
first of all, by extraction of level density and emission
probability of the nuclear reaction product from the spectra
(cross sections) of nuclear reactions.

\section{Non-removable uncertainty of experimental data analysis
of some nuclear-physics experiments}\hspace*{16pt}

By analysis of experimental data in low energy nuclear physics
(at least in some its sections) is really used the postulate on
principle possibility of unambiguous determination of desired
nuclear parameters. For example, level density in fixed interval
$\Delta E$ for given nuclear excitation energy and emission
probability of some nuclear reaction product at their de-excitation.
Or excitation -- at decay of higher-lying levels. However,
the experience of determination of $\rho$ and radiative strength
functions $k$ from intensities of two-step gamma-cascades
\cite{Meth1,PEPAN-2005} together with analysis of possibilities
of existing methods of analogous experiments \cite {PHAN7210}
shows that their unambiguous determination is impossible.
It is true, at least, for the present and for region of high
level density. Practically, it follows from this circumstance that the $\rho$
and $k$ values can be determined only with inevitable systematical
error or there can be found only final interval of values of these
parameters which contains desired parameter. And its asymptotical
width is not equal to zero.

The task under consideration obviously belongs to the same class.
Id est:

(a) the parameter $\nu$ of $\chi^2$-distribution can be unambiguously
determined (with precision up to experimental uncertainty and
statistical fluctuations), but there cannot be tested all the necessary
conditions of applicability of hypothesis \cite{PT}, or

(b) there can be found only asymptotically non-zero interval of
values of parameters of expression (1). Below is realized only the
second possibility.  The more detailed description the analysis
method of and results of its test are presented in \cite{AMSa}.    

\section{Results of analysis}\hspace*{16pt}

Comparison of experimental cumulative sums with approximating
curves for 15 sets of $s$- and three sets of $p$-resonances in variants:

the distributions $K=1$ and 
superposition of four possible distributions ($K=4$) is presented
in figures 1 and 2.

\begin{figure}\begin{center}

\vspace{3cm}
\leavevmode
\epsfxsize=15cm

\epsfbox{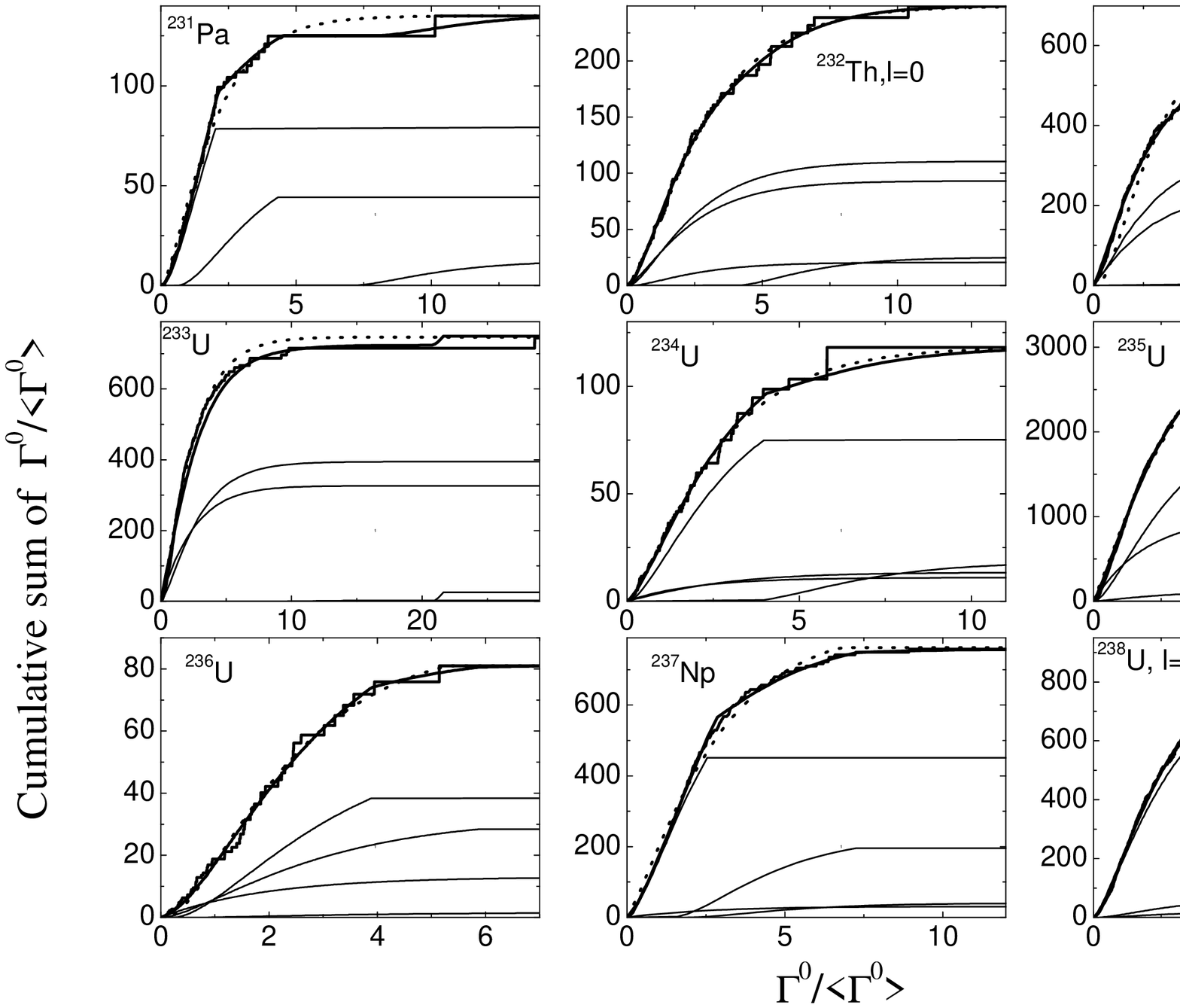} 

\end{center}
\vspace{-5.5cm}

Fig. 1. Histogram  - cumulative sum of 
$\Gamma^0_n$/$<\Gamma^0_n>$ for their values, less than given magnitude.
Thick solid curve -- the best approximation by four distributions,
dotted curve -- by one distribution for nuclei with mass  $231 \leq A \leq 238$.
Thin curves -- the most probable values of approximating functions
for $1 < K \leq 4$.
\end{figure}

\begin{figure}\begin{center}
\vspace{4cm}
\leavevmode
\epsfxsize=15cm

\epsfbox{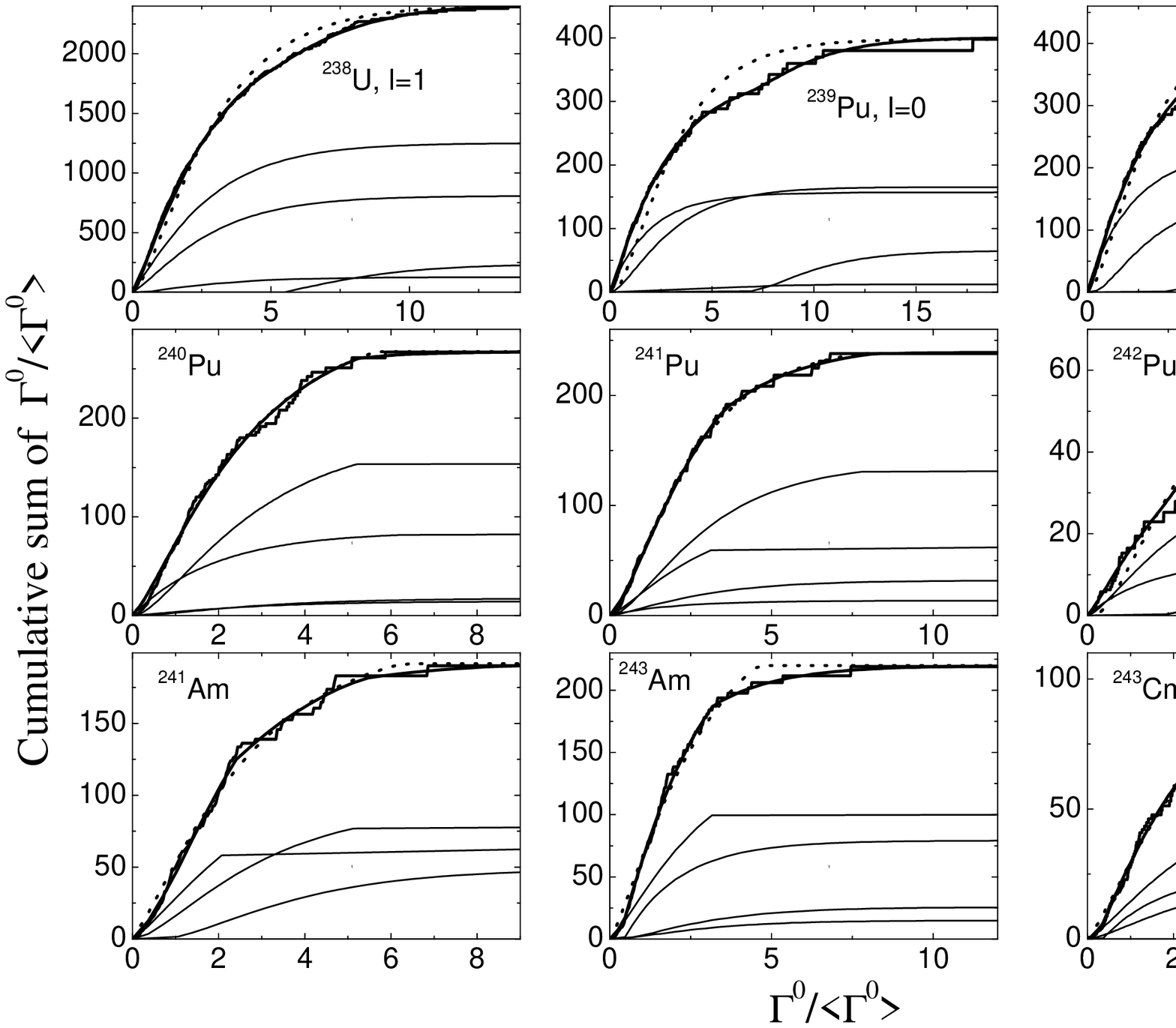} 
\vspace{-5cm}

Fig. 2. The same, as in Fig. 1, for $238 \leq A \leq 243$.
\end{center}
\end{figure}

The ratios of $\chi^2$ in function of nuclear mass for two variants
of approximation are shown in Fig. 3. (Approximation in all the cases
was performed, as a minimum, over $X$=1000 values.

It follows from these figures that approximation of the experimental
data by superposition of four distributions improved precision for the
greater part of experimental data. And maximally -- for $s$-resonances
of $A$-odd nuclei and all sets of available $p$-resonances.

The simplest possible explanation is obvious: the $2g\Gamma^0_n$ values
for different spins of resonances differ by parameters of distributions
of their neutron amplitudes. It is enough for their appearance in the
obtained experimental data as a superposition of
two distributions with different $\sigma$ and/or $b$ values.
But, the tendency of change of ratio of approximation parameter $\chi^2$
for different nuclei allows and possibility of change in shape of
width distribution when nucleus mass changes.

Large dispersion of random values $X$ brings to large fluctuations
of cumulative sums of both experimental data and model distributions
\cite{AMSa}. And, respectively, to essential variations of the best
values of parameters (1). That is why, the conclusions about possible
deviations of $b$ and $\sigma$ parameters from expected values 0 and 1,
respectively, can have, as it was mentioned above, only probabilistic
character. In Fig. 4 are compared frequency distributions of these
parameters for modeling sets with N=150, 500 and 2000 random $X$ values.
Modeling was performed for the variant of absence of omission of small
$X$ values and for omission corresponding to exclusion of $L=30\%$ of
their lowest magnitudes (linearly changing with number of random value).

\begin{figure}\begin{center}

\vspace{2cm}
\leavevmode
\epsfxsize=9cm

\epsfbox{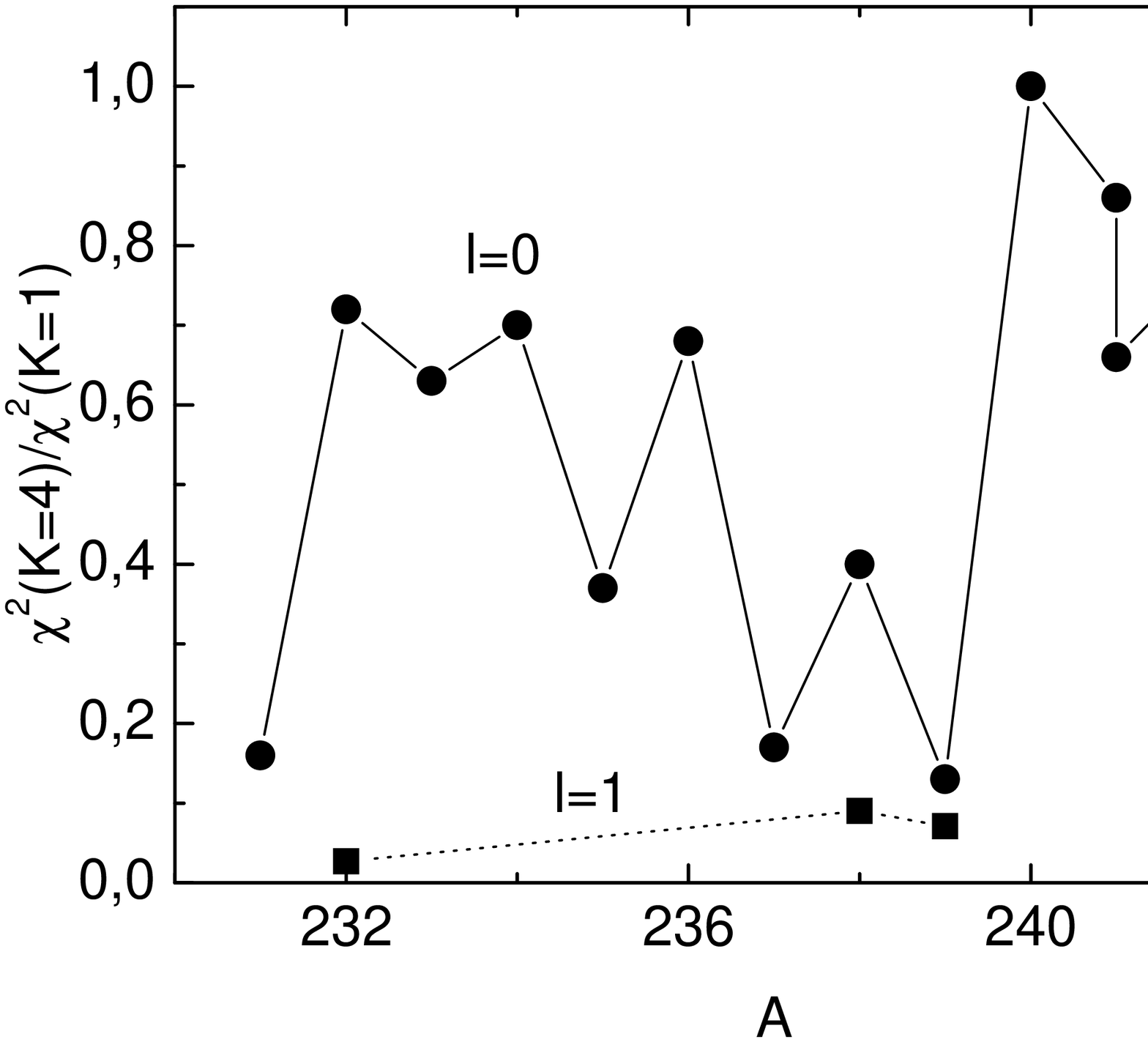} 
\end{center}
\vspace{-3cm}

Fig. 3.  The ratio of the lowest  $\chi^2$ values for the sets
from $K$ approximating distributions for actinides under consideration.

\end{figure}

\begin{figure}[htbp]
\begin{center}

\vspace{3cm}
\leavevmode
\epsfxsize=12cm

\epsfbox{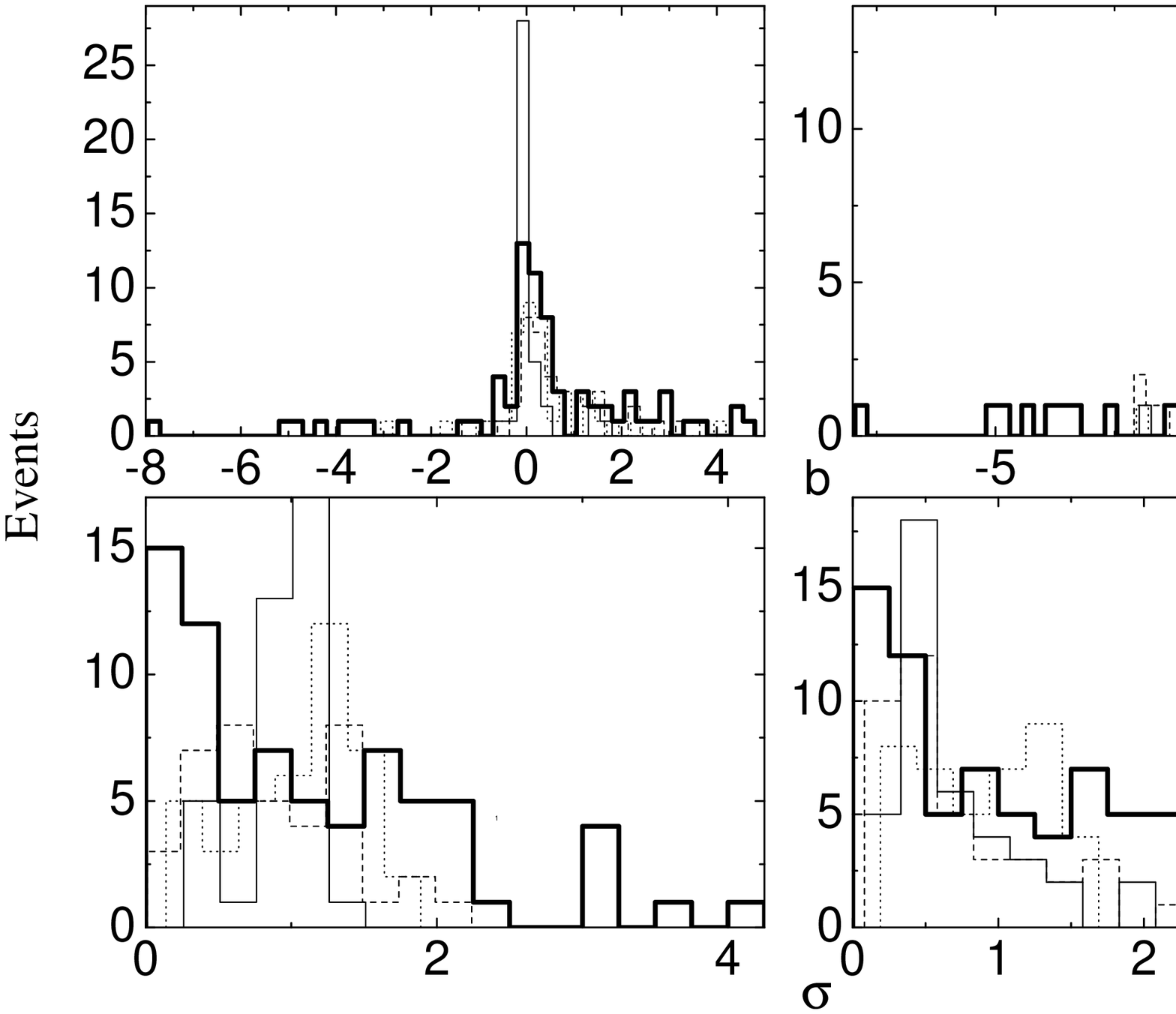}
\end{center}
\vspace{-4cm}

Fig. 4. The comparison of the frequency distributions of
appearance of given values of $b$ (upper) and $\sigma$ (lower) row
accordingly for $K = 1$. Left column -- in modeling are
included all the possible random values; right column -- there are
excluded $L=30\%$ of the lowest random values in each tested set.
Thick solid curve -- experimental data set, thin solid, dashed
and dotted curves -- the data for $N$=2000, 500 and 150 random
values in modeled sets.

\end{figure}

\begin{figure}\begin{center}

\vspace{3cm}
\leavevmode
\epsfxsize=12cm

\epsfbox{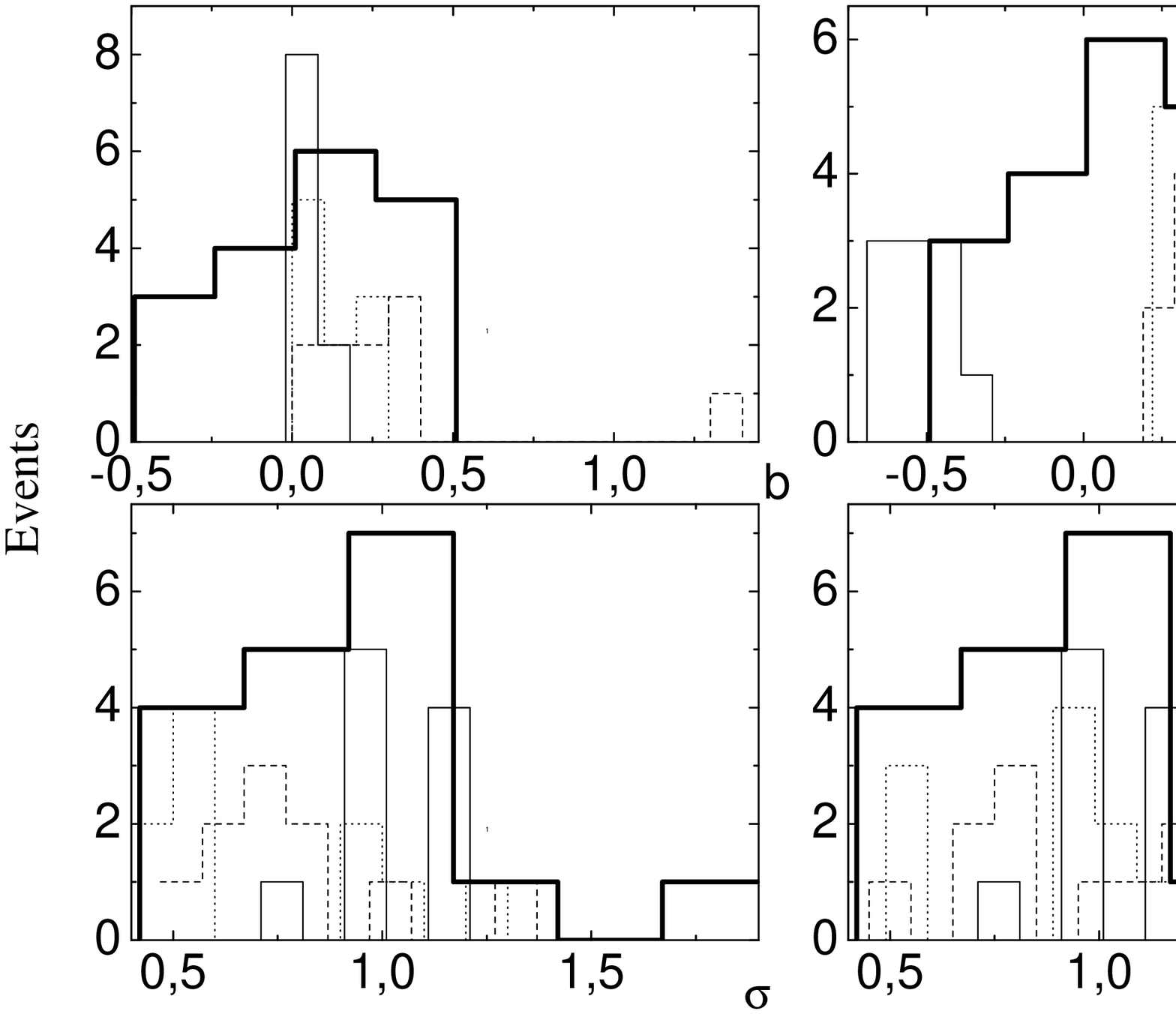}  
\end{center}
\vspace{-4cm}

Fig. 5. The same, as in Fig. 4, for the case $K = 4$.
\end{figure}

The widths of corresponding distributions decrease when $N$
increases. They are minimal for $L=0$ and in all practical cases --
less than the maximal width of $b_k$ and $\sigma_k$ experimental
frequency distribution. Figures 1 and 2 permit one to conclude
that deviation of experimental width distribution from the
Porter-Thomas distribution appears itself mainly at
$X=(\Gamma^0_n/<\Gamma^0_n>)>2-5$. Discrepancy between experimental
data and hypothesis \cite{PT} at less values of $X$ can be related,
first of all, to omission of weak resonances or to other systematical
errors of experiment. But, it is not excluded and possibility of real
deviations of parameters $b$ and $\sigma$ from the values corresponding
to hypothesis \cite{PT}.

Although deficiency of experimental $\Gamma_n$ values for analyzed
here actinides did not allow one to get unambiguous conclusions on
real parameters of expression  (1), the information from the data of
estimation of mean spacing between their resonances is much more
 unexpected \cite{AMSc}.

\section{Conclusion}\hspace*{16pt}

1. The analysis performed shows that the probability of correspondence of
distribution $2g\Gamma^0_n$ to the unique functional dependence ($K =1$)
in nuclei of different mass is less than to the set of visibly different
functions ($K =4$). Therefore, any quantitative test of hypothesis
\cite{PT} should be performed by comparison of two or more different
model notions in maximal set of nuclei.

2. The results of performed analysis, probably, do not contradict to
hypothesis \cite{PT} on equality of mean value of amplitude to zero
for the main part of the determined $\Gamma^0_n$ values.
More precise statement on this account can be made only after
significant decrease of experimental threshold of resonance registration.

3. The more unambiguous conclusions on equality of dispersion of
experimental distribution of $\Gamma^0$ to the fitted value
cannot be made on basis of present analysis.

4. Parameters of approximation of the experimental data permit
a presence of, at least, superposition of two sets of resonances
for $K>1$  with different structure of wave functions.

5. Possible deviation of parameter $\nu$ of the Porter-Thomas
distribution from $\nu = 1$ for the majority of experimental
sets of neutron widths can be interpreted only after model
estimation of its random fluctuations.

6. Unambiguous conclusions on problems considered here require
very significant increase of sets of resonances with experimentally
determined $\Gamma^0_n$ ($\Gamma^1_n$) values at their accordingly
decreased distortions.

\end{document}